\documentstyle[emulateapj,graphics]{article}

\begin{document}
\newcommand{\eg}{{\sl e.g.}}
\newcommand{\ie}{{\sl i.e.}}
\newcommand{\etal}{{\sl et al. }}

\newcommand{\rhounit}{\mbox{$M_\odot \,$pc$^{-3}$}}
\newcommand{\vunit}{\mbox{km\,s$^{-1}$}}
\newcommand{\Me}{\mbox{$M_\oplus$}}
\newcommand{\Msun}{\mbox{$M_{\odot}$}}
\newcommand{\Rsun}{\mbox{R$_{\odot}$}}
\newcommand{\Lsun}{\mbox{L$_{\odot}$}}
\newcommand{\ltsimeq}{\raisebox{-0.6ex}{$\,\stackrel 
        {\raisebox{-.2ex}{$\textstyle <$}}{\sim}\,$}} 
\newcommand{\gtsimeq}{\raisebox{-0.6ex}{$\,\stackrel
        {\raisebox{-.2ex}{$\textstyle >$}}{\sim}\,$}} 
\newcommand{\prpsimeq}{\raisebox{-0.6ex}{$\,\stackrel
        {\raisebox{-.2ex}{$\textstyle \propto $}}{\sim}\,$}}

\newcommand{\JHK}{$J\!H\!K~$}
\newcommand{\degs}{$^{\circ}$}
\newcommand{\pms}{$\pm$}
\newcommand{\chis}{\mbox{$\chi^{2}~$}}
\newcommand{\rchis}{\mbox{$\chi^{2}_{\nu}~$}}
\newcommand{\kms}{$\,$km$\,$s$^{-1}$}
\newcommand{\gcm}{$\,$g$\,$cm$^{-3}$}
\newcommand{\ergs}{$\,$erg$\,$s$^{-1}$}


\title{Infrared Photometric Variability of GX13+1 and GX17+2}
\author{Reba M. Bandyopadhyay\altaffilmark{1}}
\vspace{0.3mm}
\affil{Department of Astrophysics, Oxford University, Keble Road, Oxford OX1 3RH, UK}
\vspace{0.3mm}
\author{Philip A. Charles}
\vspace{0.3mm}
\affil{Department of Physics \& Astronomy, University of Southampton, Southampton SO17 1BJ, UK}
\vspace{1mm}
\author{Tariq Shahbaz} 
\vspace{0.3mm}
\affil{Instituto de Astrofisica de Canarias, C/Via Lactea, s/n, 38200 La Laguna, Tenerife, Spain}
\vspace{1mm}
\author{R. Mark Wagner}
\vspace{0.3mm}
\affil{Steward Observatory, University of Arizona, Tucson, AZ 85720, USA}
\altaffiltext{1}{rmb@astro.ox.ac.uk}

\begin{abstract}

\noindent

We present infrared photometry of the Galactic Bulge X-ray binary systems GX13+1 and GX17+2 obtained in 1997 July and August using OSIRIS on the 1.8m Perkins Telescope at Lowell Observatory.  GX13+1 clearly varies over $\sim$0.6 magnitudes in the $K$-band.  Our light curve suggests a modulation on a timescale of $\sim$20 days, which is in agreement with previously proposed orbital periods for the system.  The IR counterpart of GX17+2 is also variable in the $K$-band over $\sim$0.8 magnitudes on a timescale of days to weeks, extending the variability first seen by Naylor, Charles, \& Longmore (1991).  We discuss the implications our data have for Deutsch \etal's (1999) identification of ``star A'' as the true IR counterpart of GX17+2.  The variability observed in our photometry of the blend of star A and the foreground star NP Ser implies a $\sim$4 magnitude intrinsic variation in the $K$-band for GX17+2.
\end{abstract}

\keywords{binaries: close -- infrared: stars -- X-rays: stars -- stars: individual: GX13+1, GX17+2 -- accretion, accretion discs}

\section{Introduction}
In low-mass X-ray binaries (LMXBs), mass is transferred from a late-type star to its highly compact companion, either a neutron star or a black hole, via an accretion disc.  These systems can be classified according to location within the Galaxy, accretion characteristics, and luminosity (van Paradijs \& McClintock 1995).  Mapping of the X-ray sky has revealed a dozen bright X-ray sources within $15^{\circ}$ longitude and $2^{\circ}$ latitude of the Galactic Centre (Warwick \etal 1988).  Known as the ``galactic bulge'' or ``bright bulge'' sources (henceforth referred to as ``GBS''), these LMXBs are among the most luminous X-ray sources in the Galaxy (typical $L_{X} \sim 10^{38}$ \ergs).  However, the GBS remain the most poorly understood group of LMXBs, due to the heavy obscuration in the direction of the Galactic bulge which makes optical study nearly impossible.  The GBS are thought to be neutron star LMXBs, and quasi-periodic oscillations (QPOs) have been detected in several systems (\eg van der Klis 1989).  Attempts to detect orbital variability in the GBS have generally been unsuccessful, suggesting that their periods may be longer than those of canonical LMXBs (Charles \& Naylor 1992, hereafter CN92).  On the basis of their X-ray colour-colour diagrams, neutron star LMXBs have also been divided into two classes, known as $Z$ and atoll sources (see van der Klis 1995 for a review).  In this scheme, six of the brightest neutron star LMXBs (which includes several of the GBS) are classified as $Z$ sources, while the remainder fall into the atoll category.  However, recently this distinction has come into question, as several atoll sources show $Z$-type colour-colour diagrams when examined on a long timescale (Muno \etal 2001, Gierlinski \& Done 2001).  Nevertheless clear differences remain between the X-ray luminosity and spectral evolution of the canonical $Z$ sources and that of the atolls.


The infrared provides us with an ideal window for observing these highly obscured systems.  Observing in the IR has two primary advantages:  the late-type secondaries in LMXBs are brighter relative to the accretion discs, and, more importantly for the GBS, the ratio of $V$- to $K$-band extinction is nearly 10 (Naylor, Charles, \& Longmore 1991, hereafter NCL91).  Over the past eight years, we have developed a program of IR observations of X-ray binaries (XRBs), beginning with the discovery via colours or variability of candidates for the IR counterparts to the X-ray sources using precise X-ray and radio locations (NCL91).  Following this photometric survey we performed a spectroscopic survey of LMXBs, obtaining IR spectra of a number of systems, including Sco X-1, GX13+1, and Sco X-2 (Bandyopadhyay \etal 1997 and 1999, hereafter B97 and B99).  The spectra enabled us to place constraints on the spectral types of the mass donors in these systems, indicating that the secondaries in the GBS may be evolved rather than main-sequence stars.


If, as the evidence suggests, the companion stars in the GBS are indeed evolved stars, we would expect any variability in the IR light curve due to the orbital period to occur on a timescale of days or weeks, rather than hours as in canonical LMXBs.  To search for long orbital periods in the atoll LMXB GX13+1 and the $Z$ LMXB GX17+2 (as classified by Hasinger \& van der Klis 1989), in 1997 we obtained IR photometry of both sources over a period of approximately 6 weeks.  An IR study of the GX13+1 field mapped by the {\it Einstein HRI} in X-rays revealed a variable IR source at the Grindlay \& Seaquist (1986) radio position of GX13+1 (CN92).  The observed variability made this IR source a very strong candidate for the counterpart to GX13+1.  The $K$-band spectrum we obtained in 1997 showed Brackett $\gamma$ emission, confirming the identification of the IR counterpart (B97).  An additional spectrum obtained in 1999 with improved S/N exhibited both Brackett $\gamma$ emission and CO absorption bands; using the technique of optimal subtraction, we determined the most likely spectral type of the secondary to be a K5 {\sc iii} (B99).  

In contrast to GX13+1, detection of the true IR counterpart to GX17+2 has proven more difficult.  GX17+2 was optically ``identified'' on positional grounds alone with a G star now known as NP Ser more than 25 years ago (Tarenghi \& Reina 1972).  However, the absence of any optical variability or spectroscopic peculiarity in a star purported to be associated with one of the most luminous X-ray sources in the Galaxy has long been a major puzzle, with most explanations requiring that the G star be simply a line-of-sight object.  Furthermore, the IR counterpart to NP Ser has shown variability, and a consistent fit for the optical and IR colours, extinction, distance, and spectral type could not be found (NCL91).  The situation has recently been clarified by HST observations of GX17+2 which resolved NP Ser into two components, the fainter of which, labelled ``star A'' by Deutsch \etal (1999), has been suggested as the true IR counterpart to GX17+2.  The unusual nature of star A has been demonstrated by Callanan \etal's (1999) Keck photometry which suggests that star A exhibits extremely large ($\sim$3.5-4 mag) IR variability.  In this paper we present long-term ($\sim$5-week) light curves of GX13+1 and GX17+2 from July and August 1997; both sources show clear variability of greater than 0.5 magnitudes in the $K$-band.

\section{Observations and Data Reduction} 

We obtained $K$-band photometry during 14 nights in 1997 between 16 July and 23 August using the Ohio State Infrared Imager/Spectrometer (OSIRIS) on the Perkins 1.8m telescope of the Ohio State and Ohio Wesleyan Universities at Lowell Observatory in Flagstaff, Arizona.  OSIRIS uses a 256$\times$256 pixel HgCdTe NICMOS III detector; see DePoy \etal (1993) for further details on the design of this instrument.  The images were taken with the f/7 camera which provides a 2.7 arcmin$^{2}$ field of view at a resolution of $0.63''$ per pixel.  The summer monsoon season affected the observing run, causing some variability in the data quality; however, typical images were obtained with $\ltsimeq 2''$ seeing.  On each night, a series of five consecutive 60 second (2 second on-chip integrations $\times$ 30 coadds) exposures were obtained for each target, with offsets of $\sim20''$ between each exposure.  The position of the object on the array was moved between exposures so that the group could later be median-stacked to produce a sky frame.  

The images were processed by first running a simple interpolation program to remove bad pixels.  A median-combined sky image was then created from the five exposures and subtracted from each image.  The resulting sky-subtracted frames were flat-fielded with an image constructed from the difference of dome flats obtained with the flat-field lamps on and off.  The data were then analyzed using the IRAF reduction task ``apphot'' with a 4.5 pixel aperture which was chosen to maximize the target signal relative to the sky background.  Relative photometry was performed using several field standards.  Finally, absolute photometry of the local standards was performed.  A journal of the IR observations with the calculated $K$ magnitudes of GX13+1 and GX17+2 is presented in Table 1.

\begin{table*}
\caption{1997 $K$ magnitudes of GX13+1 and GX17+2}
\begin{center}
\begin{tabular}{lcc}\hline
UT Date		& GX13+1 		& GX17+2 	 \\\hline 
16 July 	& 12.12$\pm$0.03 	& 14.63$\pm$0.25     \\ 
		& 11.87$\pm$0.03	& 14.79$\pm$0.32    \\
17 July 	& 12.06$\pm$0.03 	&          	    	\\ 
18 July		& 12.23$\pm$0.04 	& 14.69$\pm$0.29     \\ 
		& 12.30$\pm$0.04	&		    \\
19 July		& 12.30$\pm$0.06 	& 14.96$\pm$0.38     \\ 
		& 12.60$\pm$0.05	& 15.27$\pm$0.43    \\
		& 12.45$\pm$0.04	&		    \\
23 July		& 12.47$\pm$0.03 	& 14.48$\pm$0.14     \\ 
		& 12.46$\pm$0.04	& 14.26$\pm$0.16    \\
24 July		& 12.40$\pm$0.03 	& 14.22$\pm$0.14    \\
25 July		& 12.46$\pm$0.03 	& 14.02$\pm$0.12     \\ 
26 July		& 12.30$\pm$0.05 	& 14.26$\pm$0.28     \\ 
		& 12.24$\pm$0.03	& 14.07$\pm$0.17    \\
1 August	& 11.98$\pm$0.02	& 14.30$\pm$0.20     \\ 
12 August	& 12.22$\pm$0.03 	& 14.83$\pm$0.26     \\ 
		& 12.11$\pm$0.03	& 14.81$\pm$0.31    \\
		& 12.21$\pm$0.03	& 14.87$\pm$0.27    \\
13 August	&  			& 14.08$\pm$0.26     \\ 
14 August	& 12.54$\pm$0.04 	& 14.63$\pm$0.26    \\
20 August	& 12.10$\pm$0.03 	& 14.14$\pm$0.10    \\ 
23 August	& 11.97$\pm$0.02 	&           	     \\ \hline
\end{tabular}
\end{center}
\end{table*}	

\section{Results}

\subsection{GX13+1}

Our $K$-band light curve of GX13+1 is presented in Figure 1; these data constitute the longest IR light curve of this source published to date.  Substantial variability of $\sim$0.6 mag is clearly present over the six weeks of observation.  Visual inspection of the GX13+1 light curve suggests a characteristic variability timescale of order $\sim$20 d.  Unfortunately, the IR data are not sufficiently well sampled, nor is the observational baseline long enough, for a formal statistical analysis to yield meaningful results as to the presence of $\sim$20 d periodicities.  The lower panel of Figure 1 shows the ASM 1-day average X-ray light curve of GX13+1 during the interval of our observational program.  

The most striking feature of the CN92 light curve (which spanned 10 days) was an increase in brightness of $\sim$0.4 mag over two days (from $K \sim$ 12.3 to 11.9).  Subsequent nights did not show similar variability, and monitoring of GX13+1 over a period of 5 hrs did not show any modulation at a level greater than 0.1 mag.  Groot \etal (1996; hereafter G96) obtained Gunn {\it z} band (0.9-1.0 $\mu$m) photometry of GX13+1 over a period of 18 days, and found clear variability on a timescale of $\sim$10 days; using a Lomb-Scargle periodigram they estimate the maximum power to be at 12.6 days, with an estimated error of $\sim$1 day.  However, they note that the time interval covered by their observations was too short to formally establish a periodicity in the system.  Analysis of 1 year of RXTE ASM monitoring of GX13+1 revealed a possible 25.2 d modulation in the X-ray light curve (Corbet 1996).  In contrast to the CN92 and G96 observations, Wachter (1996; hereafter W96) found a 19.5 hr modulation from 4 nights of CTIO $K$-band photometry.  They also report the detection of the $\sim$0.4 mag variability seen by CN92, and suggest that this variation occurs on a much shorter timescale than the 12.5 or 25.2 d periods which had been suggested by G96 and Corbet. 

To search for evidence of our $\sim$20 day modulation in the X-ray light curve of GX13+1, we re-analyzed the data from the RXTE ASM.  The ASM light curve now spans approximately 5 years, in contrast to the $\sim$1 year available for Corbet's 1996 analysis.  Using the publically available ``definitive'' ASM dwell data from the RXTE archive, we used the ``perdgrm'' task in FTOOLS v5.0.1 to perform a power spectral analysis on the GX13+1 data.  The maximum peak is found at $\sim$26.7 days; however, this peak is not stable.  When the X-ray light curve is divided into 5 $\sim$1-year intervals, the frequency of the maximum peak varies between $\sim$24-45 days.  To illustrate the significance and variability of this modulation, we present a 2-D ``dynamical'' power spectrum of the GX13+1 ASM data in Figure 2.  This power spectrum is constructed from the 1-day average light curve, summing the 3 energy channels for an energy range of 1.5-12 keV.  In order to search for any periodic phenomena which might be transitory in nature, the $\sim$1700 days of data were divided into 200 day subsets, but with a spacing between them of only 100 days (i.e. overlapping the subsets each side).  Each subset was subjected to a Lomb-Scargle periodogram analysis covering the period range 4-50 days, and the resulting periodogram highlights any changes in temporal behaviour.  Figure 2 shows that variations on timescales of 20--30 days are present in the power spectra of GX13+1.  Specifically, during the intervals of MJD 51300-51400 and 51700-51800 significant peaks (contour levels $geq$12, corresponding to a 99\% confidence level) are seen at frequencies of $\sim$27 and $\sim$22 days respectively.  However, these are of a quasi-periodic nature and no truly periodic signal is consistently present at this frequency for the duration of the ASM dataset.  It seems likely that the 25.2 d variability reported by Corbet (1996) from the first year of RXTE monitoring of GX13+1 is also part of this long-timescale quasi-periodicity.







On the basis of the currently available IR photometry, we are not able to identify the $\sim$20 d variability as an orbital modulation.  A period of this length might also be associated with the precession of a warped accretion disc, with the true orbital period being substantially shorter.  Cyg X-2 is an example of this scenario; it has a 9.8 d orbital period, determined from optical spectroscopy (Casares, Charles, \& Kuulkers 1998), and a 78 d period determined from X-ray light curves and attributed to disc precession (Wijnands, Kuulkers, \& Smale 1996).  If the ratio $P_{disc}/P_{orb}$ is similar for GX13+1, we might expect an orbital period of $\sim$2.5 days for GX13+1 for a 20 d disc precession period.

However, $K$-band spectroscopy has determined the most likely spectral type of the secondary in GX13+1 to be a K5{\sc iii} (B99).  This spectral type correlates well with a 20 d orbital period for GX13+1 and rules out a period of either $\sim$2.5 days or the 19.5 hr period proposed by W96.  It therefore seems more likely that the observed $\sim$20 d variability may be an orbital modulation rather than being caused by a precessing disc.  We note that the large IR variability and the low $L_{X}/L_{opt}$ ratio in GX13+1 indicates that the disc is large and that X-ray heating of the disc and secondary contributes significantly to the IR flux from the system (CN92).  This conclusion was supported by the IR spectroscopy, which indicated that the K5{\sc iii} secondary probably contributes less than half of the $K$-band flux.  We also note that the IR and X-ray light curves shown in Figure 1 appear generally uncorrelated on a timescale of $\sim$weeks, although the sampling of the IR light curve is insufficient for a formal correlation analysis.  Therefore we suggest that the IR variability of GX13+1 arises from a combination of causes, with short-term IR emission from the X-ray heating superimposed on an underlying long orbital modulation.  This scenario would account for both the long-term ($\geq$10 days) variability shown in Figure 1 and the shorter timescale variability seen by CN92 and W96, while simultaneously being consistent with the classification of the companion as a K5{\sc iii} star ($M \simeq 5\Msun, R \simeq25\Rsun$).




\subsection{GX17+2}

We observed the field of the GBS GX17+2 on 13 nights (see Table 1) in 1997; the $K$-band light curve is presented in Figure 3.  Variability over $\sim$0.8 mag is clearly present over the six weeks of observation.  Our $K$-band light curve again confirms the variability seen by NCL91, and extends it to show that the IR source varies over $\sim$0.6-0.7 mag on a timescale of $\sim$10 days.  However, our data are not sufficiently well-sampled to determine a periodicity in the IR photometric modulation.  To search for an X-ray period of comparable length, we performed a power spectral analysis of the RXTE ASM dwell data of GX17+2.  Similarly to our analysis for GX13+1, in Figure 4 we present a dynamical power spectrum of the GX17+2 ASM 1-day average light curve.  We find no significant peak below $\sim$200 days.  

Using HST IR observations of the GX17+2 field, Deutsch \etal (1999; hereafter D99) were able to identify a new candidate counterpart to the X-ray source, located 0.9'' north of NP Ser.  They found a photometric magnitude of $H \approx$ 19.8 and an $H$-band reddening of 2 for this new candidate, labelled ``star A''.  Subsequently, Callanan \etal (1999) obtained a Keck $K$-band image of GX17+2 and were able to resolve NP Ser and star A, deriving $K =$ 14.5 and 14.9 respectively.  Archival $K$-band images from 1998 examined by Anderson \etal (1999) also showed star A to be only slightly fainter than NP Ser.  However, the $H$-band magnitude and reddening found by D99 implies $K\sim$18.5 for star A at the time of their observations.  Taken together, these observations suggest that star A exhibits $\sim$3.5-4 magnitudes of variability in the $K$ band.  This extreme modulation is consistent with the prediction of D99 that the optical/IR counterpart to GX17+2 should be highly variable.  

Our $K$-band images obtained with the 1.8m telescope at Lowell are of insufficient resolution to resolve NP Ser and star A.  However, the blended light curve of the two stars, shown in Figure 3, does exhibit significant variability.  Assuming that at the peak of its brightness, star A has $K \sim$ 14.9, and that at minimum $K \sim 18.5$, we can estimate the effect of the variability of star A on the blended light curve of the two stars.  At star A's minimum, we expect to see flux {\it only} from NP Ser at its presumably constant magnitude, $K \approx$14.5.  When star A is at its maximum, we observe the combined flux of {\it both} NP Ser and star A, which are then of roughly similar magnitude.  At the maximum of GX17+2, then, we would expect to see an increase in the blended light curve of 0.75 magnitudes from its minimum of 14.5.  This estimate is clearly consistent with the modulation shown in our GX17+2 light curve, which varies by $\sim$0.8 magnitudes.  Note that the minimum values we observed for the NP Ser/star A blend are consistent within the errors with $K \sim$14.5, the magnitude of NP Ser.  Therefore the photometric modulation seen in our blended light curve of NP Ser and star A is almost certainly a result of the large variability intrinsic to star A.  The combination of the revised positional estimate for GX17+2 and the observed extreme photometric variability of star A are strong evidence for the identification of star A as the IR counterpart to GX17+2.  We note that the ASM light curve of GX17+2 (shown in the lower panel of Figure 3) is remarkable for its constancy compared to the dramatic IR variability during this same interval, similar to the behaviour seen in GX13+1 (Figure 1).

\section{Conclusions}

Our $K$-band photometry of GX13+1 shows evidence for a $\sim$20 d modulation.  We note that a 20 d period correlates extremely well with that expected for a Roche-lobe filling binary with a K5{\sc iii} secondary (B99).  On the basis of the observed variability in the IR together with the identification of the secondary's spectral type, we suggest that the observed $\sim$20 d modulation may be the orbital period of GX13+1.  We note that to conclusively identify any period of order $\sim$20 d in GX13+1, it will be necessary to obtain a well-sampled IR light curve with a baseline of $\gtsimeq$60 d.  We also find evidence in the GX13+1 ASM X-ray light curve for quasi-periodicity on a timescale of 20-30 days, but this modulation is not consistently present throughout the $\sim$5-year dataset.

We also observe a $\sim$10 d modulation in the IR light curve of GX17+2.  Our photometry was performed on the blend of the foreground star NP Ser and the recently identified counterpart candidate star A; the observed variability is consistent with that expected as a result of a large modulation of the fainter of the two stars.  The IR photometric modulation and revised position for GX17+2 strongly supports the identification of star A as the true counterpart to the X-ray source.  Previous attempts at obtaining an IR spectrum of GX17+2 have been unsuccessful, no doubt as a result of the small separation between NP Ser and star A as well as star A's large photometric variability which would make it nearly undetectable for a significant fraction of its IR period.  However, now that these difficulties are recognized, future IR spectroscopic observations of star A will hopefully prove more successful in investigating the physical nature of the extraordinary variability of the true IR counterpart.




\section{Acknowledgements}

The data reduction was carried out using the $\sc iraf$ and $\sc ark$ software packages at the Oxford University Starlink node.  A portion of this work was performed while RMB was supported by a National Research Council Research Associateship at the Naval Research Laboratory.


\clearpage
\newpage
\clearpage

\onecolumn
\begin{figure} 
\plotone{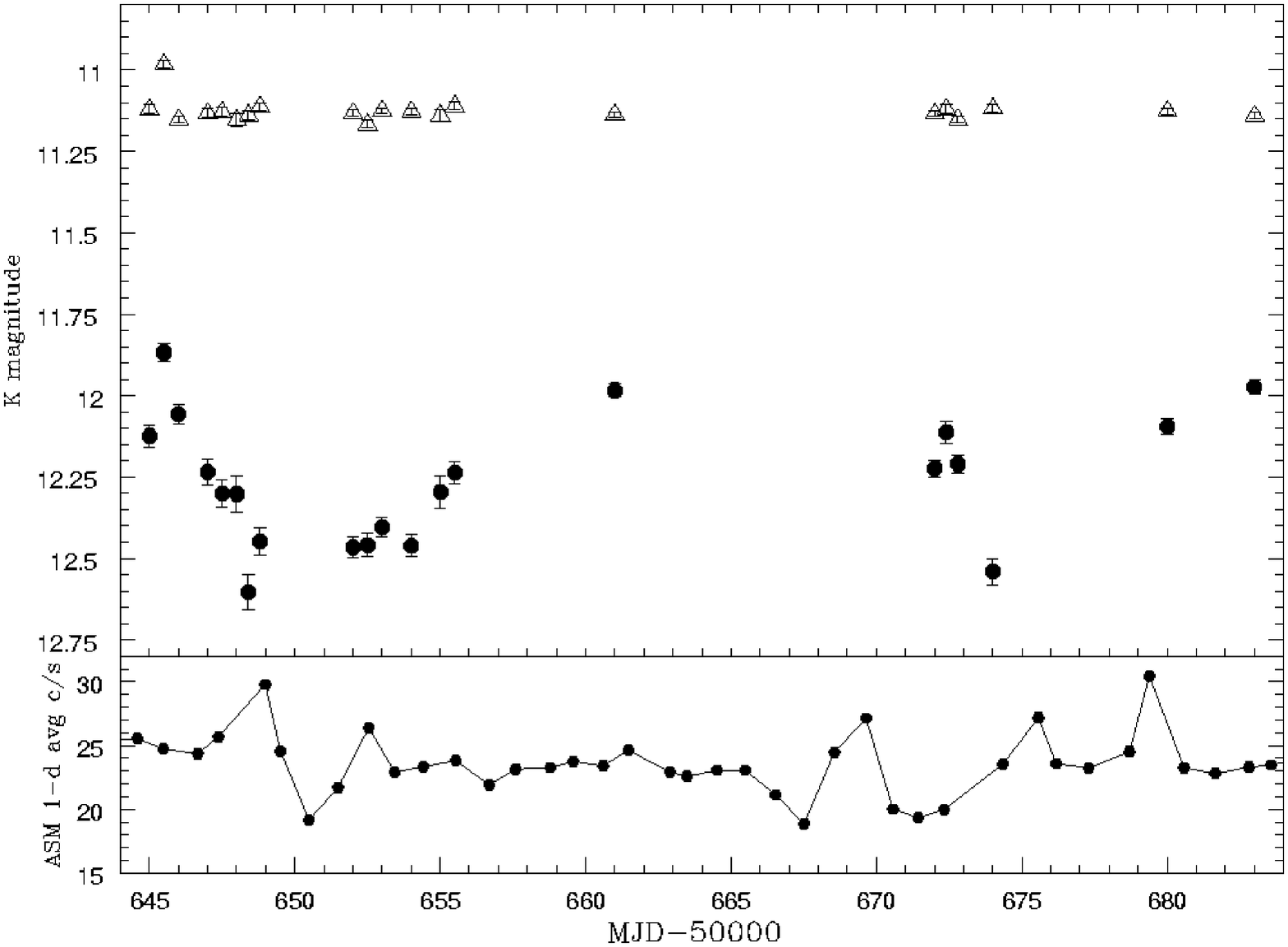}
\caption{{\it Upper panel:} $K$-band relative photometry of GX13+1 (filled circles) from July-August 1997.  A local standard is plotted for comparison (open triangles).  Note the clear variability of GX13+1 suggesting a modulation on the order of $\sim$20 days.  Typical photometric errors for GX13+1 are $\pm$0.03 mag. {\it Lower panel:} Simultaneous {\it RXTE} ASM 1-day average X-ray light curve of GX13+1.}
\end{figure} 


\begin{figure} 
\plotone{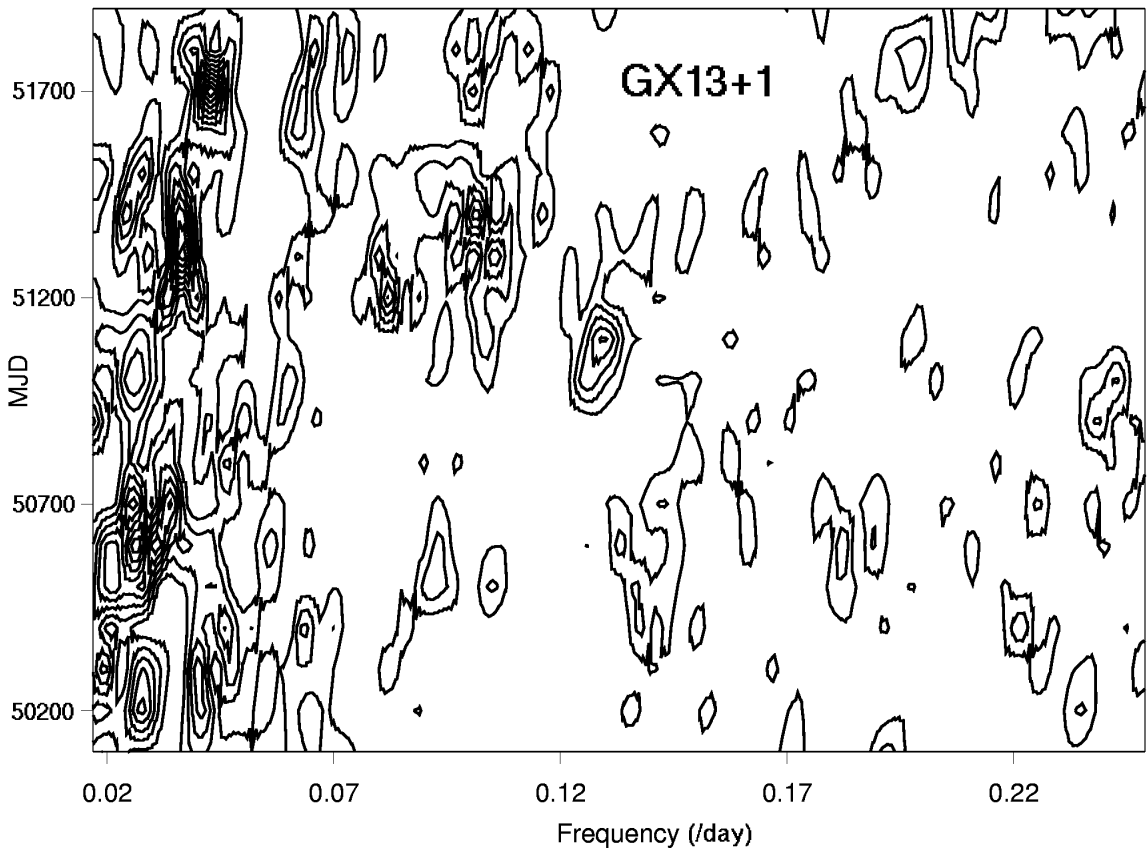}
\onecolumn{\caption{Dynamical power spectrum of the GX13+1 RXTE ASM X-ray light curve.  Note the significant, variable peaks around MJD 51300-51400 and MJD 51700-51800 (contour levels $>$10) at frequencies of $\sim$27 and $\sim$22 days respectively.}}
\end{figure} 


\begin{figure} 
\plotone{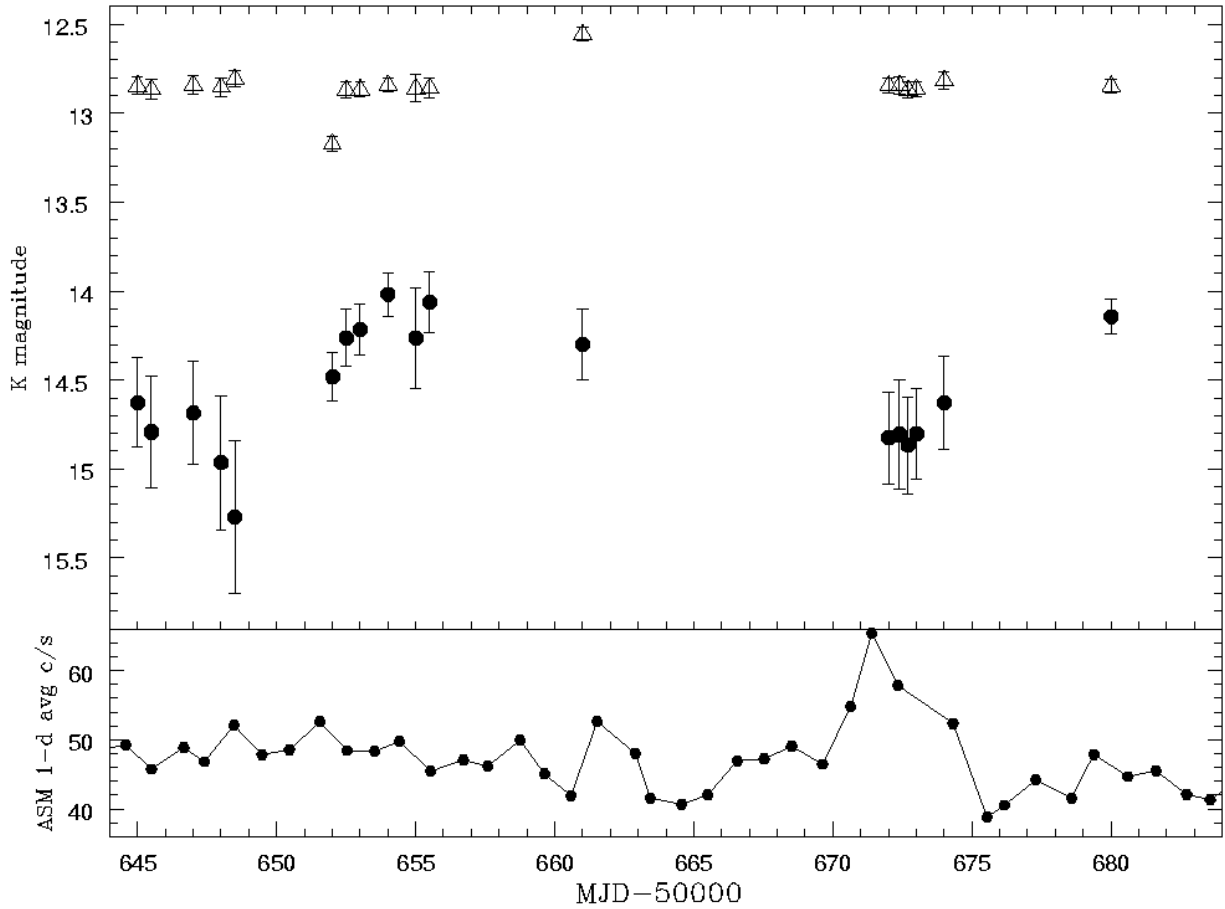}
\caption{{\it Upper panel:} $K$-band relative photometry of the NP Ser/star A blend, the stars at the position of GX17+2 (filled circles).  A local standard is plotted for comparison (open triangles).  Despite the large errors (generally $\pm$0.3 mag), the blend of NP Ser and star A does exhibit significant variability of order $\sim$0.8 mag. {\it Lower panel:} Simultaneous {\it RXTE} ASM 1-day average X-ray light curve of GX17+2. \label{fig2}}
\end{figure} 

\begin{figure} 
\plotone{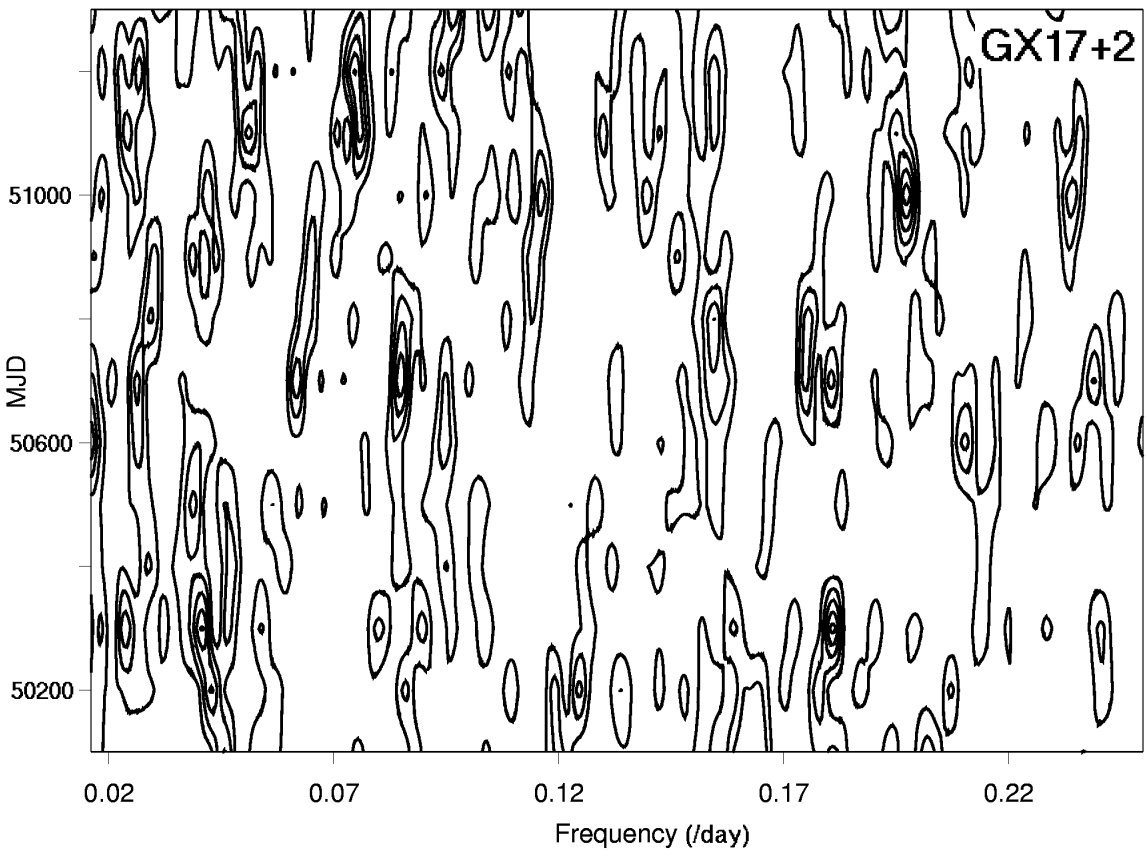}
\onecolumn{\caption{Dynamical power spectrum of the GX17+2 RXTE ASM X-ray light curve.  The highest contour level is $<$8, indicating that no highly significant peaks were found.}}
\end{figure} 

\end{document}